\documentstyle[amsfonts,multicol,fancyheadings,eqsecnum,prb,aps]%
{revtex}

\begin{document}
\draft

\title{Gauge Transformations in Einstein-Yang-Mills Theories} 

\author{J.\ M.\ Pons \cite{pons}}
\address{Departament d'Estructura i Constituents de la 
Mat\`eria, Universitat de Barcelona, \\
and Institut de F\'\i sica d'Altes Energies,\\ 
Diagonal 647, E-08028 Barcelona, Catalonia, Spain} 

\author{D.\ C.\ Salisbury \cite{salisbury}} 
\address{Department of Physics, Austin College, 
Sherman, Texas 75090-4440, USA}

\author{L.\ C.\ Shepley \cite{shepley}}
\address{Center for Relativity, Physics Department, \\ 
The University of Texas, Austin, Texas 78712-1081, USA \\ 
~ } 

\date{Version of 22 December 1999\ --- to be submitted to 
\textit{J.\ Math.\ Phys.}}
\maketitle
\begin{abstract}

We discuss the relation between spacetime diffeomorphisms and gauge
transformations in theories of the Yang-Mills type coupled with
Einstein's General Relativity.  We show that local symmetries of the
Hamiltonian and Lagrangian formalisms of these generally covariant
gauge systems are equivalent when gauge transformations are required
to induce transformations which are projectable under the Legendre
map.  Although pure Yang-Mills gauge transformations are projectable
by themselves, diffeomorphisms are not.  Instead the projectable
symmetry group arises from infinitesimal diffeomorphism-inducing
transformations which must depend on the lapse function and shift
vector of the spacetime metric plus associated gauge transformations. 
Our results are generalizations of earlier results by ourselves and by
Salisbury and Sundermeyer.

~
\end{abstract}

\pacs{04.20.Fy, 04.40.-b, 11.10.Ef, 11.15.-q \hfill gr-qc/9912086} 

\setlength{\columnseprule}{0pt}\begin{multicols}{2}


\section{Introduction}
\label{sec:intro}

In a recent paper \cite{pons/salisbury/shepley/97} we discussed the
relation between diffeomorphisms and gauge transformations in General
Relativity.  Specifically, gauge transformations are required to be
projectable under the Legendre map, and therefore they must depend on
the lapse function and shift vector of the metric in a given
coordinate neighborhood.  Therefore, it is not the diffeomorphism
group, which acts on the underlying manifold, which is the gauge
group.  The gauge group acts on the dynamical variables; its structure
is fixed by the dynamical model; but each element may also be
interpreted as a diffeomorphism.  More precisely, each pair consisting
of an element of the gauge group and a metric on which it acts
determines a spacetime diffeomorphism (which affects tensors in the
usual way).

Here we extend the discussion to include spacetimes having a
Yang-Mills type field coupled to General Relativity.  Our work is an
extension of a more formal treatment by Pons and Shepley
\cite{pons/shepley/95}.  Some of these results were earlier obtained
by Salisbury and Sundermeyer
\cite{salisbury/sundermeyer/83a,salisbury/sundermeyer/83b} (and
others), but we feel we here have given them a broader foundation,
namely one based on projectability under the Legendre map.  In
addition all gauge variables are retained in the new treatment.

We find that pure Yang-Mills gauge transformations meet our
requirement of projectability.  Gauge transformations which act like
diffeomorphisms not only have to be coupled to the metric as in the
vacuum case but also require associated gauge transformations.

In Section \ref{sec:Yang-Mills} we briefly recount the general
treatment of diffeomorphism-invariant theories.  We discuss
Einstein-Yang-Mills field theory and describe (infinitesimal) gauge
transformations therein.  We show explicitly how these transformations
must depend on the lapse function and shift vector of the spacetime
metric and what associated Yang-Mills gauge transformations they must
have if they are to be projectable under the Legendre map.  In Section
\ref{sec:structure}, we calculate the group structure functions and
the canonical group generators.  Section \ref{sec:conclusion}
concludes with a general discussion of our results and future
extensions.  These will include the application of our procedures to
the real triad formulation \cite{goldberg/88,friedman/jack/88} and to
the Ashtekar formulation \cite{ashtekar/91} of General Relativity.


\section{Yang-Mills Theories and General Relativity} 
\label{sec:Yang-Mills}

As in our previous paper \cite{pons/salisbury/shepley/97}, following
work of Batlle {\it et al} \cite{batlle/gomis/gracia/pons/89}, we
begin with a Lagrangian $L(q, \dot q)$ which does does not depend
explicitly on $t$.  A Noether Lagrangian symmetry 
\[ \delta L = dF/dt
\]
results in an equation for
\begin{equation}
G:={\partial L\over\partial \dot q^i} \delta q^i - F\ , 
\label{G.def}
\end{equation}
namely \[ [L]_i\delta q^i + {dG\over dt} = 0\ , \] $[L]_i$ being the
Euler-Lagrange functional derivative of $L$: \[ [L]_i = \alpha_i -
W_{is}\ddot q^s\ , \] 
where \[ W_{ij}:= {\partial^2L\over\partial\dot
q^i\partial\dot q^j} \quad {\rm{}and} \quad \alpha_i:= -
{\partial^2L\over\partial\dot q^i\partial q^s}\dot q^s + {\partial
L\over\partial q^i}\ .
\]

When the mass matrix or Legendre matrix ${\bf{}W}=(W_{ij})$ is
singular, there exists a kernel for the pullback ${\cal F}\!L^*$ of
the Legendre map ${\cal F}\!L$ from configuration-velocity space $TQ$
(the tangent bundle $TQ$ of the configuration space $Q$) to phase
space $T^*\!Q$ (the cotangent bundle).  This kernel is spanned by
vector fields whose components $\gamma^i_A$ ($A$ ranges over the
number of these vectors) are a basis for the null vectors of $W_{ij}$. 
The Hamiltonian technique eases the calculation of the $\gamma^i_A$:
\begin{equation} \gamma^i_A = {\cal F}\!L^* \left(
\partial\phi_A\over\partial p_i \right) , \label{gam}
\end{equation}
where the $\phi_A$ are the Hamiltonian primary first class
constraints.  Note that these constraints are here assumed to be
effective (if not, they can be made effective; however, problems can
arise when ineffective, secondary constraints, occur
\cite{pons/salisbury/shepley/98,garcia/pons/98}).

The equation satisfied by $G$ implies
\begin{equation}
\gamma^i_A{\partial G\over\partial \dot q^{i}} = 0\ , 
\label{criterion}
\end{equation}
showing that $G$ is projectable to a function $G_{\rm H}$ in~$T^*\!Q$;
that is, it is the pullback of a function (not necessarily unique) in
$T^*\!Q$: \[ G = {\cal F}\!L^*(G_{\rm H})\ ,
\]
(first pointed out by Kamimura \cite{kamimura/82}).  The function
$G_{\rm H}$ is determined up to the addition of linear combinations of
the primary constraints, but it is in general possible to absorb them
and conclude that a projectable variation must be of the form
\begin{equation} \delta q^i = {\cal F}\!L^* \left({\partial G_{\rm H}
\over\partial p_i}\right)\ .  \label{delta}
\end{equation}
We will apply this result to diffeomorphisms and to Yang-Mills gauge
transformations below.


\subsection{Yang-Mills Gauge Transformations} 
\label{subsec:Yang-Mills.yangmills}

The Yang-Mills Lagrangian density ${\cal L}_{\rm YM}$ is a functional
of the vector potential fields $A^{i}_{\mu}$, where the internal index
$i$ ranges over $\{1,\cdots,n\}$, where $n$ is the dimension of the
gauge group, and $\mu$ is a spacetime index ($\mu=0,\cdots,3$).  (We
will be using lower-case indices from the beginning of the alphabet,
$a,b,\dots$, as spatial indices, $a,b=1,2,3$.)  The field tensor
derived from these potential fields is
\begin{equation}
F^{i}_{\alpha\beta} = A^{i}_{\beta,\alpha}-A^{i}_{\alpha,\beta} 
-C^{i}_{jk}A^{j}_{\alpha}A^{k}_{\beta} \ , \label{field}
\end{equation}
where the comma denotes partial differentiation and where $C^{i}_{jk}$
are the structure constants of the gauge group.  The Yang-Mills
Lagrangian density is given by
\begin{equation}
{\cal L}_{\rm YM} = -{1\over4}\sqrt{|^{4}g|} 
F^{i}_{\mu\nu}F^{j}_{\alpha\beta}
g^{\mu\alpha}g^{\nu\beta}C_{ij} \ ,
\label{YM.lag}
\end{equation}
where $C_{ij}$ is a nonsingular, symmetric group metric and $^{4}g$ is
the determinant of the spacetime metric tensor.  (In a semi-simple
group, $C_{ij}$ is usually taken to be $C^{s}_{it}C^{t}_{js}$; in an
Abelian group, one usually takes $C_{ij}=\delta_{ij}$.)

The derivatives of ${\cal L}_{\rm YM}$ with respect to t he velocities
of the configuration space variables, $\dot A^{i}_{\alpha}$ (here
$\dot{}$ is $\partial/\partial t$), give the tangent space functions
$\hat P^{\alpha}_{i}$ corresponding to the phase space conjugate
momenta:
\begin{equation}
\hat P^{\alpha}_{i} := {\partial{\cal L}_{\rm YM} 
\over\partial\dot A^{i}_{\alpha}}
= \sqrt{|^{4}g|}F^{j}_{\mu\nu}g^{\alpha\mu}g^{0\nu}C_{ij}\ . 
\label{mom.YM}
\end{equation}
The Legendre map ${\cal F}\!L$ is defined by mapping $\hat
P^{\alpha}_{i}$ to $P^{\alpha}_{i}$ in phase space.  Because of the
antisymmetry of the field tensor, the primary constraints are
\begin{equation} 0 = \hat P_{i} := \hat P^{0}_{i} ={\partial {\cal
L}_{\rm YM}\over\partial\dot A^{i}_{0}} =
\sqrt{|^{4}g|}F^{j}_{\mu\nu}g^{0\mu}g^{0\nu}C_{ij} \ . 
\label{prime.YM}
\end{equation}
A generator of a projectable gauge transformation thus must be
independent of $\dot A^{i}_{0}$.

An infinitesimal Yang-Mills gauge transformation is defined by an
array of gauge fields $\Lambda^{i}$ and transforms the potential by
\begin{equation} \delta_{\rm R}[\Lambda]A^{i}_{\mu} =
-\Lambda^{i}_{,\mu} - C^{i}_{jk}\Lambda^{j}A^{k}_{\mu}
\label{gaugevarA}
\end{equation}
(we use the notation $\delta_{\rm R}[\Lambda]$ for this Yang-Mills
rotation variation to distinguish it from other variations defined
later, and we write $\delta_{\rm R}$ if the $[\Lambda]$ may be
understood in context).  We denote this transformation by
\begin{equation} \delta_{\rm R}A^{i}_{\mu} =: - ({\cal
D}_{\mu}\Lambda)^{j} \ , \label{gaugevarcovA}
\end{equation}
where ${\cal D}_{\mu}$ is the Yang-Mills covariant derivative (in its
action on spacetime scalars and Yang-Mills vectors).  Under this
transformation, the field transforms as
\begin{equation}
\delta_{\rm R}F^{i}_{\mu\nu}
= -C^{i}_{jk}\Lambda^{j}F^{k}_{\mu\nu} \ , 
\label{gaugevarF}
\end{equation}
where we work to first order in $\Lambda^{i}$ and use the Jacobi
identity \[ C^{i}_{j\ell}C^{\ell}_{mn} + C^{i}_{m\ell}C^{\ell}_{nj} +
C^{i}_{n\ell}C^{\ell}_{jm} = 0 \ .
\]
The Yang-Mills Lagrangian ${\cal L}_{\rm YM}$ is invariant under this
transformation provided that the group metric obeys \[ C^{k}_{\ell
i}C_{kj} = - C^{k}_{\ell j} C_{ki} \ \] (which it will if
$C_{ij}=C^{s}_{it}C^{t}_{js}$).

The variation $\delta_{\rm R}$ is clearly independent of $\dot
A^{i}_{0}$ and so is projectable.

\subsection{Diffeomorphisms}
\label{subsec:Yang-Mills.diffeos}

The configuration space variables for General Relativity are the
components of the metric tensor
\begin{eqnarray}
ds^{2} & = & g_{\mu\nu}dx^{\mu}dx^{\nu}
\nonumber \\
& = & -N^{2}dt^{2} + g_{ab}(dx^{a}+N^{a}dt)(dx^{b}+N^{b}dt) \ , 
\label{metric}
\end{eqnarray}
where $N$ is the lapse function, $N^{a}$ the components of the shift
vector, and $g_{ab}$ is our notation for the spatial metric.  The
inverse of $g_{ab}$ is $e^{ab}$: \[ e^{ac}g_{bc} = \delta^{a}_{b} \ .
\]
We will use $g$ for the determinant of the spatial metric; the
relationship between it and the determinant of the spacetime metric is
\[ ^{4}g = -N^{2}g \ .
\]
In matrix form the metric and its inverse are: 
\begin{eqnarray*}
(g_{\mu\nu}) & = &
\pmatrix{ -N^2 + N^{c}N^{d}g_{cd} & g_{ac}N^{c} \cr 
g_{bd}N^{d} & g_{ab}} \ , \\
(g^{\mu\nu}) & = & \pmatrix{-1/N^2 & N^a/N^2 \cr 
N^b/N^2 & e^{ab} - N^aN^b/N^2} \ .
\end{eqnarray*}

The General Relativity Lagrangian density is \cite{wald/84} 
\begin{equation}
{\cal L}_{\rm GR} = N \sqrt{g} ({}^3\!R + K_{ab}K^{ab} 
- (K^{a}_{a})^2) \ ,
\label{GR-lag}
\end{equation}
where ${}^3\!R$ is the scalar curvature computed from the 3-metric
(${}^3\!R={}^3\!R_{ab}e^{ab}$, where ${}^3\!R_{ab}$ is the 3-metric
Ricci tensor) and $K_{ab}$ is the second fundamental form (extrinsic
curvature; indices raised by $e^{ab}$ or lowered by $g_{ab}$) for the
constant-time 3-surfaces:
\begin{equation}
K_{ab}={1\over 2N}(\dot g_{ab} - N_{a|b} - N_{b|a})\ , 
\end{equation}
with ${}_|$ meaning covariant differentiation with respect to the
3-metric connection.  Thus the total Lagrangian density is
\begin{equation} {\cal L} = {\cal L}_{YM} + {\cal L}_{GR}\ . 
\label{total L}
\end{equation}

Notice that the lapse $N$ and shift $N^a$ of the 4-metric all appear,
but their time-derivatives (that is, their velocities) do not.  This
is required of any diffeomorphism invariant theory.  To be
projectable, therefore, a variation must be independent of these
velocities as well as being independent of $\dot A^{i}_{0}$ in coupled
Einstein-Yang-Mills theory.

Consider now an infinitesimal diffeomorphism, which changes the
coordinates by
\begin{equation}
\delta_{\rm D}[\epsilon] x^{\mu} = - \epsilon^{\mu} 
\label{diffeo.def}
\end{equation}
(we write $\delta_{D}$ if the $[\epsilon]$ may be understood in
context).  Under this diffeomorphism, the spacetime metric transforms
as
\begin{equation}
\delta_{\rm D}g_{\mu\nu} = g_{\mu\nu,\sigma}\epsilon^{\sigma} 
+ g_{\sigma\nu}\epsilon^{\sigma}_{,\mu}
+ g_{\mu\sigma}\epsilon^{\sigma}_{,\nu} \ . \label{difvar.g}
\end{equation}
This is the Lie derivative equation.

We will show from this equation that $\delta_{\rm D}$ is not a
projectable transformation of the form of equation (\ref{delta})
unless it is made to depend on the lapse and shift variables.  We will
also show that $\delta_{\rm D}$ is not allowed to depend on the
Yang-Mills potential $A^{i}_{0}$.  Finally, we will look at the
variation of the Yang-Mills potential itself and show that if a new
variation is defined to include a gauge transformation along with each
diffeomorphism, the new variation will be projectable.  We now proceed
with these demonstrations.

Equation (\ref{difvar.g}) implies that the variations of the lapse and
shift due to a diffeomorphism are
\begin{mathletters}\label{difvar.NNa}
\begin{eqnarray}
\delta_{\rm D}N & = & \dot N \epsilon^{0} + N_{,a}\epsilon^{a} 
+ N\dot\epsilon^{0} - NN^{a}\epsilon^{0}_{,a} \ , \label{difvar.N} \\
\delta_{\rm D}N^{a} & = & \dot N^{a}\epsilon^{0} 
+ N^{a}_{,b}\epsilon^{b} + N^{a}\dot\epsilon^{0} 
-(N^{2}e^{ab}+N^{a}N^{b})\epsilon^{0}_{,b} \nonumber \\
&& \qquad + \dot\epsilon^{a} - N^{b}\epsilon^{a}_{,b} \ . 
\label{difvar.Na}
\end{eqnarray}
\end{mathletters}%
In order to eliminate the dependence on $\dot N,\dot N^{a}$ from these
variations, it is necessary that the $\epsilon^{\mu}$ depend on the
lapse and shift \cite{pons/salisbury/shepley/97}: \begin{equation}
\epsilon^{0} = {\xi^{0}\over N} \ , \ \epsilon^{a} = \xi^{a} -
{N^{a}\over N}\xi^{0}\ , \label{xi.def}
\end{equation}
where $\xi^{0},\xi^{a}$ are independent of $N,N^{a}$. Note that 
\begin{equation}
\epsilon^{\mu} = \delta^{\mu}_{a}\xi^{a} + n^{\mu}\xi^{0}\ , 
\label{xi.norm.def}
\end{equation}
where $n^{\mu}$ is the unit normal to the $t={\rm const}$ spacelike
hypersurfaces: \[ n^{0} = {1\over N}\ ,\ n^{a}= - {N^{a}\over N}\ . 
\]

Furthermore, equations (\ref{difvar.NNa}) show that $\epsilon^{\mu}$
cannot depend on $A^{i}_{0}$: Equation (\ref{difvar.N}) has a term
$N\dot\epsilon^{0}$ which would involve $\dot A^{i}_{0}$ otherwise;
and similarly, equation (\ref{difvar.Na}) has a term
$\dot\epsilon^{a}$ which would involve $\dot A^{i}_{0}$ unless such a
dependence is outlawed.

Under a diffeomorphism, the Yang-Mills potential transforms as a
covariant vector field under Lie differentiation: \begin{equation}
\delta_{\rm D}A^{i}_{\mu} = A^{i}_{\mu,\sigma}\epsilon^{\sigma} +
A^{i}_{\sigma}\epsilon^{\sigma}_{,\mu} \ .  \label{vardif.A}
\end{equation}
The variation of $A^{i}_{a}$ is clearly independent of $\dot N$, $\dot
N^{a}$, $\dot A^{i}_{0}$ and so is projectable.  However, the
$\delta_{\rm D}$ variation of $A^{i}_{0}$ is: \begin{equation}
\delta_{\rm D}A^{i}_{0} = \dot A^{i}_{0}\epsilon^{0} +
A^{i}_{0}\dot\epsilon^{0} + A^{i}_{a}\dot\epsilon^{a} +
A^{i}_{0,a}\epsilon^{a} \ .  \label{vardif.A0}
\end{equation}
It clearly is not projectable, nor does the dependence of
$\epsilon^{\mu}$ on the lapse and shift, equation (\ref{xi.def}), and
the non-dependence of $\epsilon^{\mu}$ on $A^{i}_{0}$ help.  What is
needed is a combined diffeomorphism and gauge transformation.

Therefore, to $\delta_{\rm D}$ we add a gauge transformation
$\delta_{\rm R}[M]$ defined by a gauge field $M^{i}$: 
\begin{eqnarray}
(\delta_{\rm D}+\delta_{\rm R}[M])A^{i}_{0} & = & \dot
A^{i}_{0}\epsilon^{0} + A^{i}_{0}\dot\epsilon^{0} +
A^{i}_{a}\dot\epsilon^{a} + A^{i}_{0,a}\epsilon^{a} \nonumber \\
& & - \dot M^{i} -C^{i}_{jk}M^{j}A^{k}_{0}\ . \label{vardif-M}
\end{eqnarray}
The most direct way of making this variation projectable, that is, to
cancel the first three terms on the right side, clearly is to choose
$M^{i}$ to be $A^{i}_{\sigma}\epsilon^{\sigma}$ (since the resulting
addition of a term involving $\dot A^{i}_{a}$ is harmless).  To this
expression may be added an arbitrary additional gauge transformation,
of course, provided it will not result in terms involving $\dot N,
\dot N^{a},\dot A^{i}_{0}$ in equation (\ref{vardif-M}).  The
subtraction from $A^{i}_{\sigma}\epsilon^{\sigma}$ of the expression
$A^{i}_{a}\xi^{a}$ represents just such a transformation; what remains
will be a term proportional to $n^{\mu}$, according to equation
(\ref{xi.norm.def}).  For what comes later, therefore, we find it
convenient to define $\delta_{\rm D}+\delta_{\rm R}[M]$ by using
\begin{equation} \ M^{i} := A^{i}_{\sigma}n^{\sigma}\xi^{0}\ . 
\label{M.def}
\end{equation}

To this variation may be added an arbitrary pure Yang-Mills gauge
transformation, and so a general projectable variation will depend on
the descriptors \[ \xi^{A} := (\xi^{0},\xi^{a},\Lambda^{i})\ , \]
there being $4+n$ functions in all.  In summary, a general projectable
variation $\delta$ acts as a combined infinitesimal diffeomorphism and
gauge transformation of the form:
\begin{mathletters}\label{d}
\begin{eqnarray}
\delta N & = & \dot\xi^{0} + \xi^{a}N_{,a} -N^{a}\xi^{0}_{,a} \ , 
\label{d.N} \\
\delta N^{a} & = & \dot\xi^{a}
- Ne^{ab}\xi^{0}_{,b} + N_{,b}e^{ab}\xi^{0} 
\nonumber \\
&& \qquad + N^{a}_{,b}\xi^{b}
-N^{b}\xi^{a}_{,b} \ ,
\label{d.Na} \\
\delta g_{ab} & = & \dot g_{ab}{\xi^{0}\over N} 
+g_{ab,c}\left(\xi^{c} -{N^{c}\xi^{0}\over N}\right) +g_{cb}
\left(\xi^{c}_{,a}-{N^{c}_{,a}\xi^{0}\over N}\right) 
\nonumber \\
&& \qquad +g_{ac}
\left(\xi^{c}_{,b}-{N^{c}_{,b}\xi^{0}\over N}\right)\ , 
\label{d.gab} \\
\delta A^{i}_{0} & = & A^{i}_{a}\dot\xi^{a} +A^{i}_{0,a}\xi^{a} 
+F^{i}_{0a}{N^{a}\xi^{0}\over N}
\nonumber \\
&&\qquad -\dot\Lambda^{i}-C^{i}_{jk}\Lambda^{j}A^{k}_{0}\ , 
\label{d.Ai0} \\
\delta A^{i}_{a} & = & F^{i}_{0a}{\xi^{0}\over N} 
+ F^{i}_{ab}{N^{b}\xi^{0}\over N}
+ A^{i}_{b}\xi^{b}_{,a} + A^{i}_{a,b}\xi^{b} 
\nonumber \\
&&\qquad - \Lambda_{,a} - C^{i}_{jk}\Lambda^{j}A^{k}_{a} \ . 
\label{d.Aia}
\end{eqnarray}
\end{mathletters}%


\subsection{Hamiltonian dynamics}
\label{subsec:hamiltonian}

To discuss the group structure functions and the canonical group
generators, we work in the Hamiltonian formulation.  First, consider
the Lagrangian energy for the Yang-Mills part of the action:
\begin{eqnarray} \hat{\cal H}_{\rm YM} & := & \dot A^{i}_{\alpha}\hat
P^{\alpha}_{i} - {\cal L}_{\rm YM} \nonumber \\
& = & {N\over2\sqrt{g}}C^{ij}g_{ab}\hat P^{a}_{i}\hat P^{b}_{j} 
+ N^{a}\hat P^{b}_{i}F^{i}_{ab}
\nonumber \\
& & +{N\sqrt{g}\over4}C_{ij}e^{ac}e^{bd}F^{i}_{ab}F^{j}_{cd} 
- A^{i}_{0}{\cal D}_{a}\hat P^{a}_{i} \ , \label{HhatYM}
\end{eqnarray}
where $C^{ij}$ is the matrix inverse of the group metric $C_{ij}$, and
we performed an integration by parts to obtain the last term.

Similarly, we can define the Lagrangian momentum functions for the
Hilbert action:
\begin{equation}
\hat p^{ab} := {\partial {\cal L}_{\rm GR}\over\dot g_{ab}} 
= \sqrt{g}(K^{ab} - K^{c}_{c}e^{ab}) \ , \label{momGR}
\end{equation}
and then compute the Lagrangian energy:
\begin{eqnarray}
\hat{\cal H}_{\rm GR} & := & \hat p^{ab}\dot g_{ab} \nonumber \\
& = &{N\over\sqrt{g}}\left(\hat p_{ab}\hat p^{ab} 
-(\hat p^{a}_{a})^{2}\right)
- N\sqrt{g}\,^3\!R
\nonumber \\
&& \qquad - 2N^{a}\hat p^{b}_{a|b}\ ,
\label{HhatGR}
\end{eqnarray}
where the last term results from an integration by parts. 

Thus the canonical Hamiltonian (whose pullback under the Legendre
transformations is the Lagrangian energy) is of the form
\begin{equation} H_{c} = \int d^{3}x\,N^A {\cal H}_{A} \ ,
\label{ham}
\end{equation}
where $N^A$ are the seven variables $N,N^a,-A^i_0$ whose conjugate
momenta $P_A=\{p,p_{a},-P_{i}\}=0$ are the primary constraints, and
${\cal H}_{A}=\{{\cal H}_{0},{\cal H}_{a},{\cal H}_{i}\}$.  The time
derivatives of the primary constraints are secondary constraints: \[
\dot P_A=\{P_A,H_{c}\}=-{\cal H}_A\ .  \] There are no more
constraints.  Explicitly: \begin{mathletters}
\label{secondary}
\begin{eqnarray}
{\cal H}_{0} & = & {1\over2\sqrt{g}}C^{ij}g_{ab} 
P^{a}_{i} P^{b}_{j}
+{\sqrt{g}\over4}C_{ij}e^{ac}e^{bd}
F^{i}_{ab}F^{j}_{cd}
\nonumber \\
& &\qquad +{1\over\sqrt{g}}
( p_{ab} p^{ab} - ( p^{c}_{c})^{2})
-\sqrt{g}\,^{3}\!R \ ,
\label{secondary.H0} \\
{\cal H}_{a} & = & P^{b}_{i}F^{i}_{ab}
- 2 p^{b}_{a|b} \ ,
\label{secondary.Ha} \\
{\cal H}_{i} & = & {\cal D}_{a} P^{a}_{i} \ . 
\label{secondary.Hi}
\end{eqnarray}
\end{mathletters}%
We summarize our notation in the following list: 

\noindent\begin{tabular}{lcccccccc}
&&&&&&&& \\
Configuration variables:~ & $g_{ab}~$ & $A^{i}_{a}~$ 
& $N$ & & $N^{a}$ & & $A^{i}_{0}$ & \\
Momentum variables: & $p^{ab}$ & $P^{a}_{i}$ 
& $p$ & & $p_{a}$ & & $P_{i}$ & \\
Primary constraints: & &
         & $p$ & $=$ & $p_{a}$ & $=$ & $P_{i}$ & $=0$ \\        
         Secondary constraints: & &
         & ${\cal H}_{0}$ & $=$ & ${\cal H}_{a}$ & $=$           
         & ${\cal H}_{i}$ & $=0$ \\
&&&&&&&& \\
\end{tabular}

The equations of motion which follow from the Hamiltonian equations
(\ref{ham}) are (see \cite{wald/84}):
\begin{mathletters}\label{EoM}
\begin{eqnarray}
\dot g_{ab} &=& \{g_{ab},H_{c}\}
\nonumber \\
&=& {2N\over\sqrt{g}}(p_{ab} - {1\over2}p^{c}_{c}g_{ab}) 
+ N_{a|b}+N_{b|a} \ ,
\label{EoM.gab} \\
\dot A^{i}_{a} &=& \{A^{i}_{a},H_{c}\}
\nonumber \\
&=& {N\over\sqrt{g}}C^{ij}g_{ab}P^{b}_{j} 
- N^{b}F^{i}_{ab} + {\cal D}_{a}A^{i}_{0}\ , 
\label{EoM.Aia} \\
\dot p^{ab} &=& \{p^{ab},H_{c}\}
\nonumber \\
&=& -{N\over\sqrt{g}}({}^{3}\!R^{ab}
                                - {1\over2} {}^{3}\!R e^{ab})
+ {N\over2\sqrt{g}}e^{ab} (p^{cd} p_{cd} 
- {1\over2} (p^{c}_{c})^{2})
\nonumber \\
&& - {2N\over\sqrt{g}}(p^{ac} p^{b}_{c}
- {1\over2} p^{c}_{c} p^{ab})
+ \sqrt{g} (N_{|ab} - e^{ab} N^{|c}{}_{c}) 
\nonumber \\
&& + (N^{c} p^{ab})_{|c}
-2 p^{c(a} N^{b)}{}_{|c}
\nonumber \\
&& +{N \over 2\sqrt{g}} C^{ij}
({1\over2}e^{ab}g_{cd}P^{c}_{i} P^{d}_{j} 
-P^{a}_{i} P^{b}_{j})
\nonumber \\
&& +{N \over 4} C_{ij}\sqrt{g}
\Big(2 F^{i}_{cd} F^{j}_{ef} e^{ca} e^{eb}e^{df} 
\nonumber \\
&&\qquad\qquad\qquad
-{1\over2}F^{i}_{cd}F^{j}_{ef}e^{ab}e^{ce}e^{df}\Big) \ , 
\label{EoM.pab} \\
\dot P^{a}_{i} &=& \{P^{a}_{i},H_{c}\}
\nonumber \\
&=& 2{\cal D}_{b}(N^{[b}P^{a]}_{i})
                 + {\cal D}_{b}(N\sqrt{g}C_{ij}e^{c[b}e^{a]d}                            F^{j}_{cd)}
\nonumber \\
        && + A^{m}_{0}C_{mj}C^{j}_{\ell i}C^{\ell k}P^{a}_{k} \ .
\label{EoM.Pai}
\end{eqnarray}
\end{mathletters}%
Of course, equations (\ref{EoM.gab},\ref{EoM.Aia}) are restatements of
the definition of momenta.

At this time we write down the most general projectable variation of
the configuration variables, dependent on the descriptors
$\xi^0,\xi^a,\Lambda^i$ (these are the same as equations (\ref{d}) but
in our present notation; we have also used the notation of covariant
differentiation with respect to the 3-metric connection):
\begin{mathletters}\label{project}
\begin{eqnarray}
\delta N & = & \dot\xi^{0} + \xi^{a}N_{,a} -N^{a}\xi^{0}_{,a} \ , 
\label{project.N} \\
\delta N^{a} & = & \dot\xi^{a}
- Ne^{ab}\xi^{0}_{,b} + N_{,b}e^{ab}\xi^{0} 
\nonumber \\
&& \qquad + N^{a}_{|b}\xi^{b}
-N^{b}\xi^{a}_{|b} \ ,
\label{project.Na} \\
\delta g_{ab} & = &
{2\xi^0\over\sqrt{g}}(p_{ab}
- {1\over2} p^c_cg_{ab})
+ \xi_{a|b} + \xi_{b|a} \ ,
\label{project.gab} \\
\delta A^{i}_{0} & = & A^{i}_{a}\dot\xi^{a} +A^{i}_{0,a}\xi^{a} 
+ {N^{a}\xi^{0}\over \sqrt{g}} P^{i}_{a} 
\nonumber \\
&&\qquad -\dot\Lambda^{i}-C^{i}_{jk}\Lambda^{j}A^{k}_{0}\ , 
\label{project.Ai0} \\
\delta A^i_a & = &
{\xi^0\over\sqrt{g}}C^{ij}g_{ab} P^b_j
+ A^i_b\xi^b_{|a} + A^i_{a|b}\xi^b
\nonumber \\
& & \qquad - \Lambda^i_{,a} - C^i_{jk}\Lambda^i A^k_a \ . 
\label{project.Aia}
\end{eqnarray}
\end{mathletters}%
Note also for future reference that the variations of $A_{\mu}^{i}$
which result from an infinitesimal spatial diffeomorphism
$x'^{\mu}=x^{\mu}-\delta^{\mu}_{a} \xi^{a}$ plus a gauge rotation with
descriptor $\Lambda^{i} = A^{i}_{b} \xi^{b}$ are
\begin{mathletters}\label{spacediff}
\begin{eqnarray}
\delta A^{i}_{0} &=& \xi^{a} F_{0a}^{i}
= {\xi^{a} \over \sqrt{g}} N P^{i}_{a}
+ \xi^{a} N^{b} F_{ba}^{i}\ ,
\label{spacediff.a} \\
\delta A^{i}_{a} &=& \xi^{b} F_{ab}^{i}\ . 
\label{spacediff.b}
\end{eqnarray}
\end{mathletters}%

We turn now to variations of the conjugate momenta.  Observe that
under time-foliation-altering transformations, we require their
time-derivatives.  These gauge transformations are therefore
implementable only on trajectories which are solutions of the
equations of motion.

To find the variations of $p^{ab}$, we use the fact that $p^{ab}$
appear in the four-dimensional connection coefficients
$\Gamma^{\alpha}_{\beta\gamma}$.  Thus $p^{ab}$ can be calculated from
the four-dimensional connection by
\begin{equation}
p^{ab} = {1\over N}{\cal G}^{abcd}\Gamma^{0}_{cd} \ , 
\label{momconnect}
\end{equation}
where
\begin{equation}
{\cal G}^{abcd} := \sqrt{g}(e^{ac}e^{bd} - e^{ab}e^{cd})\ . 
\label{defscrG}
\end{equation}
The inverse of this object is
\begin{equation}
{\cal G}_{abcd} = {1\over\sqrt{g}}
(g_{ac}g_{bd} - {1\over2}g_{ab}g_{cd})
\end{equation}
in the sense that
\begin{equation}
{\cal G}_{abcd}{\cal G}^{cdef}
= \delta^{e}_{a}\delta^{f}_{b} \ .
\end{equation}
The general variation of the connection coefficients (under an
infinitesimal diffeomorphism defined by
$x'^{\mu}=x^{\mu}-\epsilon^{\mu}$) is
\begin{equation}
\delta\Gamma^{\alpha}_{\beta\gamma}
= - \Gamma^{\sigma}_{\beta\gamma}\epsilon^{\alpha}_{,\sigma} 
+ \Gamma^{\alpha}_{\sigma\gamma}\epsilon^{\sigma}_{,\beta} 
+ \Gamma^{\alpha}_{\beta\sigma}\epsilon^{\sigma}_{,\gamma}
+ \epsilon^{\alpha}_{,\beta\gamma}
+ \Gamma^{\alpha}_{\beta\gamma,\sigma}\epsilon^{\sigma}\ , 
\label{varnGamma}
\end{equation}
and thus
\begin{equation}
\delta\Gamma^{0}_{cd}
= - \Gamma^{\sigma}_{cd}\epsilon^{0}_{,c} 
+ \Gamma^{0}_{\sigma c}\epsilon^{\sigma}_{,b} 
+ \Gamma^{0}_{b \sigma}\epsilon^{\sigma}_{,c} + \epsilon^{0}_{,bc}
+ \Gamma^{0}_{bc,\sigma}\epsilon^{\sigma} \ . 
\end{equation}
We therefore need the following relationships: 
\begin{mathletters}\label{Gterms}
\begin{eqnarray}
\Gamma^{0}_{cd}&=& {1\over N} {\cal G}_{cdef} p^{ef} \ , 
\label{Gterms.0cd} \\
\Gamma^{0}_{0d} &=& g^{0\mu} \Gamma_{\mu\, 0d} 
= {1\over N} N_{,d} + N^{-1} N^{e} {\cal G}_{edgh} p^{gh} \ , 
\label{Gterms.00d} \\
\Gamma^{e}_{cd} &=& -{1\over N} N^{e} {\cal G}_{cdfg} p^{fg} 
+ {}^{3}\Gamma^{e}_{cd} \ .
\label{Gterms.ecd}
\end{eqnarray}
\end{mathletters}%
The calculation is far from trivial, but the most difficult part is
made somewhat easier by defining, for any function $f$,
\begin{equation} \delta'f := f'(x') - f(x) \ \Longrightarrow\ \delta f
= \delta'f + f_{,\sigma}\epsilon^{\sigma}\ .
\label{deltaprime}
\end{equation}
By concentrating on the $\delta'$ variation for
$\epsilon^{\sigma}=n^{\sigma}\xi^{0}$, using the equation of motion
for the derivative term, and then adding the rather straightforward
calculation for $\xi^{a}$ (treating $p^{ab}$ as a tensor density), we
find
\begin{eqnarray}
\delta p^{ab}&=& -\xi^{0} \sqrt{g} ({}^{3}\!R^{ab} 
- {1\over2} {}^{3}\!R e^{ab})
\nonumber \\
&& + {1\over2} \xi^{0} \sqrt{g} e^{ab}
(p^{cd} p_{cd} - {1\over2} (p^{c}_{c})^{2}) 
\nonumber \\
&&-{2\over\sqrt{g}}\xi^{0}
(p^{ac}p^{b}_{c}-{1\over2}(p^{c}_{c})^{2}) 
\nonumber \\
&& +\sqrt{g} (e^{ac}e^{bd} \xi^{0}_{|cd} 
- e^{ab} \xi^{0|c}{}_{c})
\nonumber \\
&&+{ 1\over 2\sqrt{g}} \xi^{0}C^{ij}
({1\over2} e^{ab} g_{cd}P^{c}_{i} P^{d}_{j} 
-P^{a}_{i} P^{b}_{j})
\nonumber \\
&&+{1\over4} \xi^{0}C_{ij} \sqrt{g}
(2 F^{i}_{cd} F^{j}_{ef} e^{ca} e^{eb} e^{df} 
\nonumber \\
&&\qquad\qquad\qquad
-{1\over2}F^{i}_{cd}F^{j}_{ef}e^{ab}e^{ce}e^{df}) 
\nonumber \\
                &&+ p^{ab}\xi^{c}_{,c} - \xi^{a}_{,c}p^{cb} 
                - \xi^{b}_{,c}p^{ac} + p^{ab}_{~~,c}\xi^{c} \ . 
\label{varnp}
\end{eqnarray}
The $\xi^{0}$ part of the variation can be obtained from the equation
of motion (\ref{EoM.pab}) by replacing $N$ by $\xi^{0}$ and setting
$N^{a}=0$.

To compute variations of the $P^{a}_{i}$ in principle uses the same
method, namely by using the fact that $P^{a}_{i}$ comes from a
four-dimensional object, from equation (\ref{prime.YM}).  The result
is
\begin{equation}
        \delta P^{a}_{i}
        = {\cal D}_{b}(\xi^{0}\sqrt{g}C_{ij}e^{be}e^{ad}F^{j}_{cd}) 
        +               P^{a}_{i}\xi^{b}_{,b} - \xi^{a}_{,b}P^{b}_{i}
                        + P^{a}_{i,b}\xi^{b} \ .
                                \label{varnPai}
\end{equation}
This is actually the variation
$\delta_{D}+\delta_{R}[A_{\mu}n^{\mu}\xi^{0}]$. 


\section{Symmetry generators}
\label{sec:structure}

We now turn to the generators of the projectable variations. 
Generating functions $G$ will be of the form
\cite{pons/salisbury/shepley/97}
\begin{eqnarray}
G(t) &=& \int \, d^{3}x \, (\xi^{A} G^{(0)}_{A} 
+ \dot \xi^{A} G^{(1)}_{A} ) \nonumber \\ 
&=:& \xi^{A} G^{(0)}_{A} + \dot \xi^{A} G^{(1)}_{A}\ , 
\label{GH}
\end{eqnarray}
where we shall use a repeated index to include an integration over
space as well as a sum.  The descriptors $\xi^{A}$ are arbitrary
functions.

The functions in equation (\ref{GH}) are found using an extension of
the techniques of \cite{pons/salisbury/shepley/97}: The simplest
choice for the $G^{(1)}_{A}$ are the primary constraints $P_{A}$.  The
functions $G^{(0)}_{A}$ obey
\begin{equation}
G^{(0)}_{A} = -\{G^{(1)}_{A},{\cal H}_{A} \} + pc\ , 
\end{equation}
where $pc$ represents a sum of primary constraints.  The simplest
solution for $G^{(0)}_{A}$ results in
\begin{equation}
G[\xi] = P_{A} \dot \xi^{A} + ( {\cal H}_{A} 
+ P_{C} N^{B} {\cal C}^{C}_{AB}) \xi^{A}\ , \label{G} 
\end{equation}
where the structure functions are defined by \begin{equation}
\{ {\cal H}_{A},{\cal H}_{B} \}
        =: {\cal C}^{C}_{AB} {\cal H}_{C}\ .
\end{equation}

We shall determine the structure functions by first examining the
variations generated by the secondary constraints, equations
(\ref{secondary}).  The emphasis throughout will be on the underlying
transformation symmetry group.  For this purpose we first introduce
generators associated with our secondary constraints.  Let
\begin{mathletters}\label{ggeenn}
\begin{eqnarray}
R[\xi] &:= \int d^3 x \, \xi^i {\cal H}_{ i}\ , 
\label{ggeenn.R} \\
V[{\vec \xi}] &:= \int d^3 x \, \xi^a {\cal H}_{ a}\ , 
\label{ggeenn.V} \\
S[ \xi^{0}] &:= \int d^3 x \, \xi^0 {\cal H}_{ 0}\ . 
\label{ggeenn.S}
\end{eqnarray}
\end{mathletters}%

We find that $R[\xi]$ generates a Yang-Mills rotation, so we have, for
example,
\begin{equation}
\{ A^i_{a}, R[\xi] \} = \delta_{\rm R}[\xi] A^i_{a} \ . 
\label{YYMM.R}
\end{equation}

$V[{\vec\xi}]$ generates the spatial diffeomorphism plus gauge
rotation we employed in (\ref{spacediff}): \begin{eqnarray}
\delta_{V}[\vec\xi] A^i_{a} &= &\{ A^i_{a}, V[{\vec \xi}] \} \nonumber
\\
&=& {\cal L}_{\vec \xi}A^i_{a}
+ \delta_{R}[\xi^b A_b] A^i_{a}
\nonumber \\
&=& \xi^{b} F_{ab}^{i}\label{YYMM.V} \ , \end{eqnarray} where ${\cal
L}_{\vec \xi}$ denotes the Lie derivative.  It is convenient to define
a related generator $D[{\vec \xi}]$ which generates a pure spatial
diffeomorphism: \begin{equation} D[{\vec\xi}] := \int d^3 x \xi^a
\,{\cal G}_{a} \ ,
\label{4.6}
\end{equation}
where
\begin{equation}
{\cal G}_{a}:={\cal H}_{ a}-A_{a}^{i}{\cal H}_{i}\ . 
\label{4.6a}
\end{equation}

$S[ \xi^{0}]$ generates a space-time diffeomorphism plus a gauge
rotation (neither of which by itself is projectable).  So, for
example,
\begin{eqnarray}
\delta_{S}[\xi^{0}] A^i_{a} &=& \delta_{D}[\xi^{0}] A^i_{a} + 
\delta_{R}[\xi^0 A_\mu n^\mu] A^i_{a} \nonumber \\ 
&=& {\xi^{0}\over\sqrt{g}}C^{ij} g_{ab} P^{b}_{j}\ . 
\label{4.7}
\end{eqnarray}

It is straightforward to calculate the complete Lie algebra from the
calculable action of the infinitesimal group elements on the
generators.  (The only Poisson bracket we will not calculate in this
manner is the bracket of $S[ \xi^{0}]$ with $S[\eta^{0}]$, simply
because it would be tedious, invoking time derivatives of the
3-curvature and the extrinsic curvature.)

First, a gauge rotation of ${\cal H}_{i}$ yields 
\begin{mathletters} \label{CCOMM}
\begin{equation}
\{R[\xi],R[\eta]\}= -R[[\xi,\eta]]\ .
\label{CCOMM.RR}
\end{equation}
The remaining brackets are
\begin{eqnarray}
\{R[\xi],D[{\vec \eta}]\}&=& \int d^{3}x\,\xi^i 
{\cal L}_{\vec\eta}{\cal H}_{i}
\nonumber \\
&=&-\int d^{3}x\,({\cal L}_{\vec \eta}\xi^i) {\cal H}_{i} 
\nonumber \\
&=&-R[{\cal L}_{\vec \eta}\xi]\ ,
\label{CCOMM.RD}
\end{eqnarray}
\begin{eqnarray}
\{D[{\vec \xi}],D[{\vec \eta}]\}&=&\int d^{3}x\,\xi^a 
{\cal L}_{\vec \eta}{\cal G}_{a}
\nonumber \\
&=&- \int d^{3}x\,({\cal L}_{\vec \eta}\xi^a) {\cal G}_{a} 
\nonumber \\
&=& -D[{\cal L}_{\vec \eta}{\vec \xi}]
= D[[{\vec \xi},{\vec \eta}]]\ ,
\label{CCOMM.DD}
\end{eqnarray}
\begin{eqnarray}
\{S[{\xi^{0}}],D[{\vec \eta}]\}&=&\int d^{3}x\,\xi^0 
{\cal L}_{\vec \eta}{\cal H}_{0}
\nonumber \\
&=&-\int d^{3}x\,({\cal L}_{\vec \eta}\xi^0) {\cal H}_{0} 
\nonumber \\
&=& - S[{\cal L}_{\vec \eta}\xi^{0}]\ ,
\label{CCOMM.SD}
\end{eqnarray}
\begin{equation}
\{S[{\xi^{0}}],R[{\eta}]\} = 0\ ,
\label{CCOMM.SR}
\end{equation}
\begin{equation}
\{V[{\vec \xi}],R[{\eta}]\} = 0\ .
\label{CCOMM.VR}
\end{equation}
The last two brackets result from the fact that ${\cal H}_0$ and
${\cal G}_a$ are gauge scalars.  Finally, a direct calculation yields
\begin{equation} \{S[{\xi^{0}}],S[{\eta^{0}}]\}= V[\vec \zeta]\ ,
\label{CCOMM.SS}
\end{equation}
\end{mathletters}%
where
\begin{equation}
\zeta^a := (\xi^{0}\partial_{b}\eta^{0}
- \eta^{0}\partial_{b}\xi^{0}) e^{ab}\ . 
\label{zeta}
\end{equation}

>From these brackets we next determine the brackets among the $R$,
$V$, and $S$ generators alone.  We find
\begin{mathletters} \label{BRAC}
\begin{eqnarray}
\{V[{\vec \xi}],V[{\vec \eta}]\}&=&
\{D[{\vec \xi}]+R[\xi^a A_a],D[{\vec \eta}] +R[\eta^b A_b]\} 
\nonumber \\
&=& D[[{\vec \xi},{\vec \eta}]]
+ R[{\cal L}_{\vec \xi}(\eta^b A_b)]
\nonumber \\
&&\quad -R[{\cal L}_{\vec \eta}(\xi^a A_a)] 
-R[[\xi^a A_a,\eta^b A_b]]
\nonumber \\
&=&V[[{\vec \xi},{\vec \eta}]] +R[\xi^a \eta^b F_{ab}]\ . 
\label{BRAC.VV}
\end{eqnarray}
The remaining bracket is
\begin{eqnarray}
\{S[{\xi^{0}}],V[{\vec \eta}]\} &=&
\{S[{\xi^{0}}],D[{\vec \eta}]+R[\xi^a A_a]\} 
\nonumber \\
&=& -S[{\cal L}_{\vec \eta} \xi^{0}]
- R[\eta^{a} \delta_{S}[\xi^{0}] A_{a}]
\nonumber \\
&=& -S[{\cal L}_{\vec \eta} \xi^{0}] - R[\eta^{a} 
{\xi^{0} \over \sqrt{g}}C^{ij} g_{ab} P^{b}_{j}] \ . 
\label{BRAC.SV}
\end{eqnarray}
\end{mathletters}%

We read off the following non-vanishing structure functions from the
above brackets:
\begin{mathletters}\label{STRUC}
\begin{eqnarray}
{\cal C}^a_{0' 0''} &=& e^{a b} \big(-\delta^3 (x-x') 
\partial''_b \delta^3 (x-x'')
\nonumber \\
&&\quad\qquad
     + \delta^3 (x-x'') \partial'_b \delta^3 (x-x')\big)\ , 
\label{STRUC.a00} \\
{\cal C}^a_{b' c''} &=&-\delta^3 (x-x')
\partial''_b \delta^3 (x-x'') \delta^a_c 
\nonumber \\
&&\quad + \delta^3 (x-x'') \partial'_c
\delta^3 (x-x') \delta^a_b\ ,
\label{STRUC.abc} \\
{\cal C}^0_{0' a''} &=&\delta^3 (x-x'')
                \partial'_a \delta^3 (x-x') \ ,
\label{STRUC.00a} \\
{\cal C}^i_{j' k''} &=&-C^{i}_{jk}
                \delta^3 (x-x')\delta^3 (x-x'')\ ,
\label{STRUC.ijk} \\
{\cal C}^i_{0' a''} &=& -{1 \over \sqrt{g}}C^{ij} g_{ab} P^{b}_{j} 
\delta^3 (x-x') \delta^3 (x-x'') \ ,
\label{STRUC.i0a} \\
{\cal C}^i_{a' b''} &=& F^i_{ab} \delta^3 (x-x') \delta^3 (x-x'') \ . 
\label{STRUC.iab} \\
\end{eqnarray}
\end{mathletters}%

Referring to the structure functions derived above, we obtain the
following generators, where $G_{R}[\xi]$, $G_{V}[{\vec \eta}]$, and
$G_{S}[ \zeta^{0}]$ are respectively, the gauge, spatial
diffeomorphism plus associated gauge, and perpendicular diffeomorphism
plus associated gauge generators:
\begin{mathletters} \label{GGEN}
\begin{eqnarray}
G_{R}[\xi] &=& \int d^{3}x\, \Bigl( -P_i \dot{\xi}^{i} 
+{\cal H}_i \xi^{i} - C^{k}_{ij}\xi^i A^j_0 P_k \Bigr)\ , 
\label{GGEN.R} \\
G_{V}[{\vec\eta}] &=& \int d^{3}x\, \Bigl( P_a \dot{\eta}^{a} 
+N^b F^i_{ba} P_i \eta^a
\nonumber \\
&&\quad -{1 \over \sqrt{g}}C^{ij} g_{ab} P^{b}_{j} 
N \eta^{a} P_{i} + N_{,a} P_0 \eta^a
\nonumber \\
&&\quad +N^a_{,b} P_a \eta^b - N^b \eta^a_{,b} 
P_a+\eta^{a}{\cal H}_{a}\Bigr) \ ,
\label{GGEN.V} \\
G_{S}[\zeta^{0}] &=& \int d^{3}x\,
\Bigl( P_0 \dot{ \zeta}^0 + N_{,b} P_a \zeta^{0} e^{ab} 
\nonumber \\
&&\quad - N P_a \zeta_{,b}^{0} e^{ab} - N^a P_0 \zeta_{,a}^{0} 
\nonumber \\
&&\quad + \zeta^{0}N^{a}{1 \over \sqrt{g}}C^{ij} 
g_{ab} P^{b}_{j} P_{i} +{\zeta}^{0}{\cal H}_{0}\Bigr)\ . 
\label{GGEN.S}
\end{eqnarray}
\end{mathletters}%
These generators do indeed generate the variations of all variables. 

We close this section by noting that we should recover the canonical
Hamiltonian as the generator of a global time translation.  Let us
check to confirm that this is the case.  First we seek the descriptors
$\xi^{\mu}$ which correspond to $\epsilon^{\mu}=\delta^{\mu}_{0}$,
\begin{mathletters} \label{DESCRIP}
\begin{eqnarray}
\epsilon^{0}&=&1=n^{0}\xi^{0}=N^{-1} \xi^{0}\ , 
\label{DESCRIP.0} \\
\epsilon^{a}&=&0=\xi^{a}+n^{a} \xi^{0}
=\xi^{a}-N^{-1}N^{a}\xi^{0}\ .
\label{DESCRIP.a}
\end{eqnarray}
\end{mathletters}%
We deduce that
\begin{equation}
\xi^{0}=N\ ,\ \xi^{a}=N^{a}\ .
\label{xi000}
\end{equation}

We must bear in mind that $S[\xi^{0}]+D[\vec\xi]$ with $\xi^{\mu}$
given by (\ref{xi000}) is not yet the generator of a global time
translation because $S[N]$ generates a gauge transformation with
descriptor \[ (A^{i}_{\mu}n^{\mu})\xi^{0} = (A^{i}_{0}N^{-1} -
A^{i}_{a}N^{-1}N^{a})N = A^{i}_{0} - A^{i}_{a}N^{a} \ .
\]
Thus the generator $R[A^{i}_{0} - A^{i}_{a}N^{a}]$ must be subtracted
to obtain the Hamiltonian:
\begin{eqnarray}
&&S[N] + D[N^{a}] - R[A^{i}_{0} - A^{i}_{a}N^{a}] 
\nonumber \\
&&\quad =\int d^{3}x \,( N{\cal H}_{0}+N^{a} 
{\cal G}_{ a}-(A^{i}_{0} - N^{a}A_{a}^{i}) {\cal H}_{i} 
\nonumber \\
&&\quad =\int \, d^{3}x \,( N{\cal H}_{0}+N^{a} 
{\cal H}_{a} - A^{i}_{0}{\cal H}_{ i}) \ . \end{eqnarray}
This is precisely the canonical Hamiltonian, equation (\ref{ham})! 

It is important to point out that in this final expression the gauge
variables $N,N^{a},A^{i}_{0}$ are to be thought of as arbitrarily
chosen but explicit functions of space and time.  This object will
then generate a global time translation only on those members of
equivalence classes of solutions for which $N,N^{a},A^{i}_{0}$ happen
to have the same explicit functional forms.  On all other solutions
the corresponding variations correspond to more general diffeomorphism
and gauge transformations.

In fact, every generator $G[\xi]$ in (\ref{G}) with $\xi^{0} > 0$ may
be considered to be a Hamiltonian in the following sense: $G[\xi^{A}]
= G_{R}[\xi] + G_{V}[{\vec\xi}] + G_{S}[\xi^{0}]$ generates a global
time translation on those solutions which have \begin{mathletters}
\label{HAM.DESCRIP}
\begin{eqnarray}
N = \xi^{0}\ ,
\label{zeta.0} \\
N^{a} = \xi^{a}\ ,
\label{eta.a} \\
-A^{i}_{0} + A^{i}_{a} N^{a} = \xi^{i} \ . \label{xi.i}
\end{eqnarray}
\end{mathletters}%
We have already demonstrated this fact for the non-gauge variables,
and it is instructive to verify the claim for the gauge variables $N$,
$N^{a}$, and $A^{i}_{0}$.  Substituting (\ref{HAM.DESCRIP}) into
(\ref{project}), we have
\begin{mathletters} \label{HAM.var}
\begin{eqnarray}
\delta N &=& \dot N + N^{a} N_{,a} -N^{a} N_{,a} = \dot N\ , 
\label{dN} \\
\delta N^{a} &=& \dot N^{a} -N e^{ab} N_{,b} +N e^{ab} N_{,b}
+ N^{a}_{,b} N^{b} -N^{a}_{,b} N^{b}
\nonumber \\
&&= \dot N^{a}\ ,
\label{dN.a} \\
\delta A^{i}_{0} &=& A^{i}_{a} \dot N^{a} + A^{i}_{0,a} N^{a} 
+ {N^{a} N \over \sqrt{g}} P^{i}_{a}
\nonumber \\
&& - (-\dot A^{i}_{0} +
\dot A^{i}_{a} N^{a}+A^{i}_{a} \dot N^{a}) 
\nonumber \\
&& - C^{i}_{jk} (-A^{j}_{0} +A^{j}_{a} N^{a} ) A^{k}_{0} 
= \dot A^{i}_{0}\ .
\label{dA}
\end{eqnarray}
\end{mathletters}%


\section{Conclusion}
\label{sec:conclusion}

We have been guided by the idea that a Lagrangian formulation of
combined Yang-Mills theory and General Relativity should be equivalent
to the Hamiltonian formulation.  As in a previous paper
\cite{pons/salisbury/shepley/97} we conclude that gauge
transformations for the theory must be transformations which are
projectable under the Legendre map from configuration-velocity space
(the tangent bundle) to phase space (the cotangent bundle).

We found that the most general projectable transformation coming from
a diffeomorphism must depend on the lapse function $N$ and shift
vector $N^{a}$ of the metric and must be accompanied by a Yang-Mills
gauge transformation which also depends on these quantities and on the
time component of the Yang-Mills field, $A^{i}_{0}$.  These results
had been obtained by Salisbury and Sundermeyer
\cite{salisbury/sundermeyer/83a,salisbury/sundermeyer/83b} (and
others) but from other points of view.  For example, Salisbury and
Sundermeyer found them by a requirement on the commutator of various
variations.  We feel that our approach has several advantages: It is
more direct, and it expressly indicates the equivalence of the
Lagrangian and Hamiltonian approaches.  Note that the gauge group acts
on the dynamical variables, so that the diffeomorphism group, which
one would na\"\i{}vely think would be included, is not itself part of
the gauge group.  However, the diffeomorphism group provides the basis
for the gauge group, and in this case, we can further say that the
group acts specifically on solutions of the equations of motion (the
Einstein-Yang-Mills field equations).

Since the Einstein-Yang-Mills Lagrangian does not depend on the gauge
variable velocities $\dot N$, $\dot N^{a}$, and $\dot A^{i}_{0}$,
under the Legendre map from configuration-velocity to phase space the
submanifold coordinatized by these variables is mapped to a single
point in phase space.  Thus functions on configuration-velocity space
can be the pull-back of functions on phase space only if they are
constant on this submanifold.  In particular, symmetry variation
functions on the tangent space are projectable if and only if they do
not depend on these velocities.  In this manner we have determined the
diffeomorphism and gauge variations which are projectable under the
Legendre map.

Spatial diffeomorphisms are projectable, but four-dimensional
diffeomorphisms which alter the time foliation are not.  As in the
case of pure conventional gravity the full four-dimensional gauge
group must be reinterpreted as a transformation group on the space of
metric solutions, and the group elements contain a compulsory
dependence on the lapse and shift.  We have found that in
Einstein-Yang-Mills theories even this alteration is not sufficient. 
A Yang-Mills gauge transformation which is itself dependent on the
full four-dimensional Yang-Mills connection must be added to the
diffeomorphism.  The resulting transformation group must therefore be
interpreted as a transformation group on the space of metric and
connection solutions.

It would seem straightforward to apply our ideas in other contexts,
for example in other formulations of General Relativity.  For example,
the Ashtekar formulation \cite{ashtekar/91} has many similarities to a
Yang-Mills theory.  However, it uses a complex Lagrangian and complex
Hamiltonian, and so reality conditions must be imposed.  The stability
of these conditions under the evolution governed by a complex
Hamiltonian makes the study of gauge transformations more difficult
and more interesting.  Other approaches to General Relativity also
rely on structures, such as a tetrad or a $3+1$ decomposition using
triads for the spatial metric, which are added to the metric
variables.  They, too, present added difficulties --- and interest ---
for the transformation law for the triads under diffeomorphisms must
take into account the decomposition.

We anticipate that the resulting recovery, and significant
enlargement, of the gauge symmetry group in Einstein-Yang-Mills
theories will provide insights to efforts to quantize these models. 
Future work will deal with somewhat more complicated vacuum models in
which auxiliary gravitational variables exhibit additional gauge
symmetry.  The first is a real tetrad formulation of Einstein's
general relativity \cite{pons/salisbury/shepley/98a}.  Then we shall
explore the symmetry structure of Ashtekar's complex formulation of
general relativity \cite{ashtekar/91,pons/salisbury/shepley/98b}.  The
former is actually a special case of the latter, and both feature in
recent attempts to construct a quantum theory of gravity.  Since
foliation altering diffeomorphisms and time evolution are in a sense
identical, as we have explained in this paper, we may acquire insights
into strategies for imposing the scalar constraint in quantum gravity.


\section*{Acknowledgments}

JMP and DCS would like to thank the Center for Relativity of The
University of Texas at Austin for its hospitality.  JMP acknowledges
support by CIRIT and CICYT contracts AEN95-0590 and GRQ 93-1047 and
wishes to thank the Comissionat per a Universitats i Recerca de la
Generalitat de Catalunya for a grant.  DCS acknowledges support by
National Science Foundation Grant PHY94-13063.




\end{multicols}

\end{document}